\documentclass[conference]{IEEEtran}

\usepackage{graphicx,cite,epsfig,amssymb,amsmath}
\usepackage{algorithm}
\usepackage{algorithmicx}
\usepackage{algpseudocode}

\normalsize

\usepackage{graphicx,epsfig,epstopdf,amsmath,amssymb,enumerate} 
\usepackage{cite}  
\usepackage{algpseudocode}

\ifCLASSINFOpdf
\else
\fi

\begin{document}

\topskip 20pt 
\title{A Low-Complexity Antenna-Layout-Aware Spatial Covariance Matrix Estimation Method}
\author{Shangbin~Wu$^1$ and Xiaoqing~Zhang$^2$ \\
\small{$^1$Samsung R\&D Institute UK, Communications House, Staines-upon-Thames, TW18 4QE, United Kingdom.} \\
\small{$^2$School of Information Science and Engineering, Shandong University, Jinan, 250199 China.}\\
Email: shangbin.wu@samsung.com, xiaoqing.sdu@hotmail.com

}

%

%
%

\markboth{IEEE Transactions on Vehicular Technology, vol. xx, no. xx, month 2018}%
{Submitted paper}
%



\maketitle
\vspace{-2cm}
\begin{abstract}

This paper proposed a low-complexity antenna-layout-aware (ALA) covariance matrix estimation method. In the estimation process, antenna layout is assumed known at the estimator. Using this information, the estimator finds  antenna pairs with statistically equivalent covariance values and sets their covariance values to the average of covariance values of all these antenna pairs. ALA for both uniform linear array (ULA) and uniform planar array (UPA) is discussed.  This method takes the benefit that covariance matrices do not have full degrees of freedom. Then, the proposed ALA covariance matrix method is applied to a multi-cell network. Simulations have demonstrated that the proposed method can provide better performance than the widely used viaQ method, with respect to mean square errors and downlink spectral efficiencies.
\end{abstract}

\begin{IEEEkeywords}

Covariance matrix estimation, channel estimation, massive multiple-input multiple-output.
\end{IEEEkeywords}

%
\IEEEpeerreviewmaketitle

\section{Introduction}

\IEEEPARstart{A}{s} a promising technique for the fifth generation (5G) cellular network, massive multiple-input multiple-output (MIMO), a MIMO system with tens or hundreds of antennas, has attracted attentions from both the academia and industry \cite{Andrews14}\cite{Zaidi17}. Although massive MIMO is able to maximize its usage of spatial resources due to its high spatial resolution, previous studies suggested that pilot contamination would be a key limiting factor to massive MIMO networks \cite{Marzetta10}, i.e., the spectral efficiency in a massive MIMO network will be bounded even if the number of antenna grows to infinity. 

However, recent research progresses in massive MIMO \cite{Bjornson18} have shown that massive MIMO has unlimited capacity as long as the channels satisfy a certain condition, i.e., the covariance matrices of user equipments (UEs) using the same pilot are asymptotically linearly independent. In \cite{Bjornson18}, the authors claimed that this condition can be satisfied in ordinary systems. Then,
 a multi-cell minimum mean squared error (M-MMSE) receiver was developed  and it was shown that M-MMSE can achieve unbounded spectral efficiency.

Additionally, the impact of imperfect covariance matrix in massive MIMO was discussed in \cite{Bjornson16}, where a widely used covariance matrix estimation method known as the viaQ method was proposed. The viaQ method estimated the covariance matrix by calculating the weighted average of the sample covariance matrix and its diagonal. However, the viaQ method requires the true knowledge of the covariance matrix during estimation. Authors in \cite{Neumann18} proposed a two-step procedure to reconstruct the covariance matrix. However, this procedure requires proper pilot allocation to UEs, which increases its complexity. In the latest 5G systems, in addition to uniform linear arrays (ULAs), uniform planar arrays (UPAs) are used as well \cite{TS38901}. Performance of these covariance matrix methods for UPAs is unanswered.

Therefore, to fill these research gaps, a novel low-complexity and practical covariance matrix estimation method named the antenna-layout-aware (ALA) method is proposed in the paper. This method requires the BS to have access to the layout of its antenna array only, which can be conveniently set when a BS is deployed. The contributions of this paper are summarized as follows.
\begin{enumerate}
\item This proposed method has been shown effective for both ULA and UPA layouts, which is compatible to most 5G systems.
\item Most importantly, the proposed method does not require the true knowledge of the covariance matrix.
\end{enumerate}

The rest of this paper is organized as follows. Section~\ref{sec_System_Model} gives a general description of the system model. The proposed ALA method for both ULA and UPA is detailed in Section~\ref{sec_ALA}. Simulation results and analysis are presented in Section \ref{sec_results_section}. Conclusions are drawn in Section~\ref{sec_conclusion_section}.


\section{System Model} \label{sec_System_Model}
\begin{figure} 
\centering\includegraphics[width=3.5in]{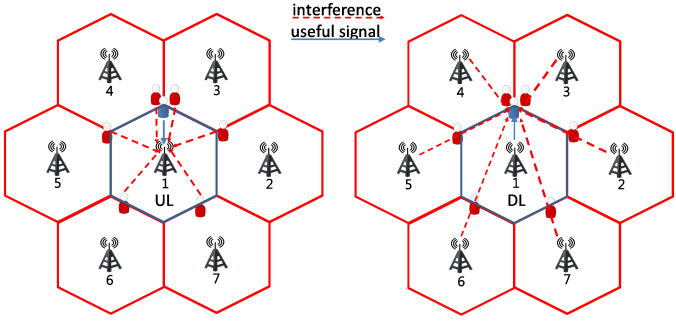} 
\caption{System diagrams of UL (left) and DL (right).}
\label{fig_System_diagram}
\end{figure}

This paper considers a mobile network with $L$ base stations (BSs) in a time-division duplexing (TDD) mode, i.e., channel reciprocity is assumed. Each BS is equipped with $N_\mathrm{t}$ antennas and corresponds to a cell. Also, there is one single-antenna UE in each cell. All UEs are assumed to share the same pilot symbols and time-frequency resources. UEs are arranged in a way such that all UEs are at the cell edge of the center cell. Moreover, the target UE lies in the line connecting an interfering UE and a neighbor BS. These settings are  as shown in Fig.~\ref{fig_System_diagram}. As a result, the worst scenario, i.e., BSs experiencing the largest interference in uplink and UEs experiencing the largest interference in downlink, is considered in this paper. 

Let $\mathbf{g}_{lk}$ be the channel vector between the $l$ BS  and the $k$th UE ($1\leq l,k \leq L$) with $\mathbf{g}_{lk}\sim\mathcal{CN}(\mathbf{0},\mathbf{R}_{lk})$, where $\mathbf{R}_{lk}$ is the corresponding spatial covariance matrix. Considering the Kronecker channel model and UPAs, $\mathbf{R}_{lk}$ can be presented as 
\begin{align}
\mathbf{R}_{lk}=\mathbf{R}^{\mathrm{v}}_{lk}\otimes\mathbf{R}^{\mathrm{h}}_{lk}
\end{align}
where $\otimes$ denotes the Kronecker product, $\mathbf{R}^{\mathrm{v}}_{lk}$ is the elevation covariance matrix, and $\mathbf{R}^{\mathrm{h}}_{lk}$ is the azimuth covariance matrix. The exponential spatial correlation model \cite{Bjornson18}\cite{Loyka01} is used in this paper. Therefore, the element in the $m$th row and the $n$th column of $\mathbf{R}^{\mathrm{h}}_{lk}$ is shown as
\begin{align}
[\mathbf{R}^{\mathrm{h(v)}}_{lk}]_{mn}= r_{\mathrm{h(v)}}^{\vert n-m\vert }e^{j(n-m)\theta_{lk,\mathrm{h(v)}}}
\end{align}
where $ r_{\mathrm{h(v)}} \in[0,1]$ is the correlation factor and $\theta_{lk,\mathrm{h(v)}}$ is the angle of arrival (AoA) in the azimuth (elevation). In the special case of a ULA, $\mathbf{R}_{lk}^{\mathrm{v}}$ equals $1$.

Assuming that the signal-to-noise ratio (SNR) between the $l$th BS and the $k$th UE is $\rho^{\mathrm{UL}}_{lk}$, the decorrelated received signal vector $\mathbf{y}_{l}$ is presented as
\begin{align}
\mathbf{y}_l={\sum_{k=1}^{L}{\sqrt{\rho^{\mathrm{UL}}_{lk}}\mathbf{g}_{lk}+\mathbf{n}^{\mathrm{UL}}}}
\label{equ_rec}
\end{align}
where $\mathbf{n}^{\mathrm{UL}}$ is the Gaussian noise in the uplink following $\mathcal{CN}(\mathbf{0},\mathbf{I})$. Then, assuming that both channel estimation and data transmission phases are within the coherence time of the channel, the minimum mean squared error (MMSE) estimate $\hat{\mathbf{g}}_{lk}$ of the channel vector ${\mathbf{g}}_{lk}$ with ideal covariance matrix can be expressed as \cite{Kailath00}
\begin{align}
\hat{\mathbf{g}}_{lk}=\mathbf{R}_{lk}\mathbf{Q}_{lk}^{-1}\left(\sum_{k=1}^{L}\mathbf{g}_{lk}+\frac{1}{\sqrt{\rho^{\mathrm{UL}}_{lk}}}\mathbf{n}^{\mathrm{UL}}\right)
\label{equ_est_channel_vector}
\end{align}
where  $\mathbf{Q}_{lk}$ is the sum of covariance matrices of users using the same pilot and a scaled identity matrix, i.e.,
\begin{align}
\mathbf{Q}_{lk}=\sum_{k=1}^L{\mathbf{R}_{lk}}+\frac{1}{\rho^{\mathrm{UL}}_{lk}}\mathbf{I}.
\end{align}
Using the orthogonality property of MMSE, the estimated channel vector $\hat{\mathbf{g}}_{lik}$ is distributed as $\hat{\mathbf{g}}_{lik}\sim \mathcal{CN}(0,\ \mathbf{\Phi}_{lk})$, with covariance matrix
\begin{align}
\mathbf{\Phi}_{lk}= \mathbf{R}_{lk}\mathbf{Q}_{lk}^{-1}\mathbf{R}_{lk}.
\end{align}

In practice, the assumption that BSs have knowledge of ideal covariance matrices is not realistic. The estimated covariance matrices $\hat{\mathbf{R}}_{lk}$ and $\hat{\mathbf{Q}}_{lk}$ can only be obtained via processing the sample covariance matrices $f\left( \hat{{\mathbf{R}}}^{\mathrm{(sample)}}_{lk} \right) $ and $f\left( \hat{{\mathbf{Q}}}^{\mathrm{(sample)}}_{lk} \right) $. The sample covariance matrices are calculated by
\begin{align}
{\hat{\mathbf{Q}}}^{\mathrm{(sample)}}_{lk}=\frac{1}{N_\mathrm{p}}\mathbf{y}_l\mathbf{y}^{\mathrm{H}}_l
\end{align}
and
\begin{align}
{\hat{\mathbf{R}}}^{\mathrm{(sample)}}_{lk}={\hat{\mathbf{Q}}}^\mathrm{sample}_{lk}-\frac{1}{N_\mathrm{p}}\mathbf{y'}_l\mathbf{y'}^{\mathrm{H}}_l
\end{align} 
where $N_\mathrm{p}$ is the number of pilot symbols and $\mathbf{y'}_l$ is the received signal excluding the associated UE's contribution. This can be done by using two slots for channel estimation. In the first slot, all UEs in the network will transmit pilot symbols simultaneously. In the second slot, only UEs in the neighboring cells will transmit pilot symbols. The processing function $f\left(\cdot \right)$ is an algorithm operating on the sample covariance matrix, which will be the viaQ method \cite{Bjornson16} or the ALA method in this paper. 

After estimating the channel in the uplink, a BS will transmit information to its UE in downlink. Meanwhile, a UE will receive both the useful signal from its associated BS and interference from neighboring BSs. Let $\rho^{\mathrm{DL}}_{lk}$ denote the downlink SNR between the $l$th BS and the $k$th UE, the received signal $z_{k}$ of the $k$th UE can be expressed as
\begin{align}
{z}_{k}=\sum_{l=1}^{L}{\sqrt{\rho^{\mathrm{DL}}_{lk}}\mathbf{g}_{lk}^\mathrm{H}\mathbf{w}_l{{s}_l}}+n^\mathrm{DL}
\label{equ_dt}
\end{align}
where $n^\mathrm{DL}$ is the zero mean unit variance Gaussian noise and $\mathbf{w}_l$ is the precoding vector of the ${l}$th BS.

Here, we follow the downlink precoder design in \cite{Jose11}, where the sum of the signal detection error and signal leakage is minimized, i.e.,
\begin{align}
\min_{\mathbf{w}_l,\alpha_l}\mathbb {E} & \Big[\|\alpha_l(\mathbf{g}^\mathrm{H}_{ll}\mathbf{w}_l{s}_l+{n}^{\rm{DL}})-{s}_l\|^2  +\sum_{q\ne l}{\| \alpha_l\mathbf{g}^\mathrm{H}_{lq}\mathbf{w}_l{s}_l\|^2}\Big|\mathbf{\hat{g}}_{lq} \Big]\nonumber
\\
&\mathrm{s.t.}\quad  \mathbf{w}_l^\mathrm{H} \mathbf{w}_l=1.
\label{equ_opt}
\end{align}
This optimization can be solved in closed form as in \cite{Jose11} and the  optimal precoding vector $\mathbf{w}_l^{\mathrm{opt}}$ can be computed as
\begin{align}
\mathbf{w}_l^{\mathrm{opt}}=\frac{1}{\alpha_l^{\mathrm{opt}}}\Big({\mathbf{\hat{g}}_{ll}^\mathrm{H}\mathbf{\hat{g}}_{ll}}+{\sum_{k=1}^L{\mathbf{R}_{lk}-\mathbf{\Phi}_{lk}}}+\mathbf{I}\Big)^{-1}\mathbf{\hat{g}}_{ll}^\mathrm{H}
\end{align}
where $\alpha_l^{\mathrm{opt}}$ is a normalization factor for the precoding vector. The downlink signal-to-noise-plus-interference ratio (SINR) $\gamma_{k}^{\mathrm{DL}}$ of the ${k}$th UE with the optimal precoding vector can be expressed as
\begin{align}
\gamma_{k}^{\mathrm{DL}}={| {\mathbb{E}}[{\mathbf{g}}_{kk}^{{\mathrm H}}{\bf w}^{\mathrm{opt}}_{k}]|^{2}\over {1\over \rho^{\rm DL}_{kk}}+{\rm var}[{\bf g}_{kk}^{{\rm H}}{\bf w}^{\mathrm{opt}}_{k}]+\sum\limits_{k'\neq k}{\mathbb{E}} \frac{\rho^{\rm DL}_{lk'}}{\rho^{\rm DL}_{lk}}\left[\left\vert {\bf g}_{lk'}^{{\mathrm{H}}}{\bf w}^{\mathrm{opt}}_{k}\right\vert ^{2}\right]}
\label{equ_sinr}
\end{align}
where $\mathrm{var}[\cdot]$ is the variance operator. It should be mentioned that the effects of non-ideal channel estimation and spatial covariance matrix estimation have been factored in (\ref{equ_sinr}).
\section{ALA Covariance Matrix Estimation} \label{sec_ALA}
After obtaining $ {\hat{\mathbf{R}}}^{\mathrm{(sample)}}_{lk}  $ and $ {\hat{\mathbf{Q}}}^{\mathrm{(sample)}}_{lk}  $, further processing can be performed to calculate the final estimated covariance matrices $ {\hat{\mathbf{R}}}_{lk}  $ and $\hat{ {\mathbf{Q}}}_{lk}  $ to improve the channel estimation performance. The viaQ method proposed in \cite{Bjornson16}\cite{Shariati14} computes $\hat{ {\mathbf{Q}}}_{lk}  $  as a weighted average of $ {\hat{\mathbf{Q}}}^{\mathrm{(sample)}}_{lk}  $ and its diagonal, i.e.,
\begin{align}
\hat{ {\mathbf{Q}}}_{lk}=(1-\kappa){\hat{\mathbf{Q}}}^{\mathrm{(sample)}}_{lk}+\kappa {\hat{\mathbf{Q}}}^{\mathrm{(sample)}}_{lk,\mathrm{diagonal}}.   
\label{equ_viaQ_regularization}
\end{align}
It was shown in \cite{Bjornson16} and \cite{Shariati14} that the optimal weight (regularization factor) $\kappa$ can be computed in closed form. However, during the calculation of the optimal weight, true knowledge of $\mathbf{Q}$ and $\mathbf{R}$ is required by the BS. This requirement is not practical.

On the contrary, the ALA method proposed in this paper does not require the true knowledge of any covariance matrices. The only additional information needed in the estimation process is the antenna layout. This can conveniently be accessed when a BS is deployed. Moveover, unlike the viaQ method, which assumes full degrees of freedom in the estimation process, the proposed ALA method takes the advantage that a covariance matrix does not have full degrees of freedom. This means that if certain antenna pairs whose relative positions are the same, then their covariance values are statistically equivalent.

As both ULA and UPA are the most widely used antenna layouts, this paper focuses on discussing the ALA covariance matrix estimation for these two layouts.
\subsection{ULA}
For a $N_\mathrm{t}$-antenna ULA, its antennas are placed in a horizontal line. Moreover, since the antenna elements are equally spaced, the covariance matrix of a ULA is a Toeplitz matrix. In this case, elements in an off-diagonal are statistically equivalent. Let us consider two pairs of antennas, PAIR($p,q$) and PAIR($p',q'$) for instance, if the two inter-element spacings are the same ($d_{pq}=d_{p'q'}$), the two covariance values will be equal, i.e., $\left[ \mathbf{R}\right]_{pq}=\left[ \mathbf{R}\right]_{p'q'}$. Therefore, all elements in the covariance matrix $ \mathbf{R}$ satisfying the above condition can be replaced by the average of them. Using the Toeplitz structure of a ULA covariance matrix, this is equivalent to computing the average value along off-diagonal lines, i.e.,

\begin{align}
\left[\hat{\mathbf{Q}} \right]_{pq}=\sum_{\overset{m,n}{m-n=p-q}}{\left[\hat{\mathbf{Q}}^{\text{(sample)}}\right]_{mn}}/\big( M-(m-n) \big)
\end{align}

Pseudo codes of the ALA covariance matrix estimation for ULA are shown in Fig.~\ref{fig_ALA_1D}. The same procedure can be used to estimate $\hat{\mathbf{R}}$.

%
    
    \begin{figure}
\hrulefill
\begin{algorithmic}[1]
\Require $\hat{\mathbf{Q}}^{\text{(sample)}}$

                \For{$p=0:N_\mathrm{t}-1$}
                \For {$q=0:N_\mathrm{t}-1$}
                        \State $\hat{\mathbf{Q}}(p,q)=\text{mean}(\text{diag}(\hat{\mathbf{Q}}^{\text{(sample)}},q-p))$
                \EndFor
                \EndFor

\end{algorithmic}

\hrulefill
\caption{ALA covariance matrix estimation for ULA.}
\label{fig_ALA_1D}
\end{figure}

\subsection{UPA}
\begin{figure}
\centering\includegraphics[width=3.5in]{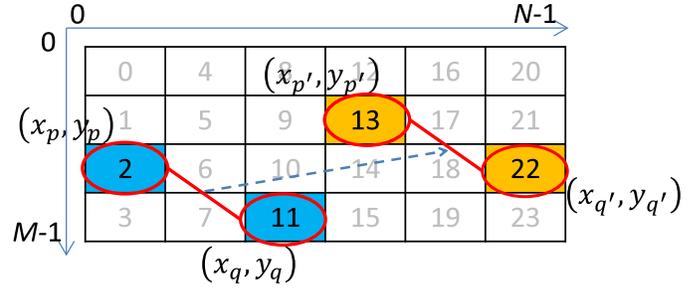} 
\caption{UPA antenna indexing and example of antenna pairs with statistically equivalent covariance values.}
\label{fig_Antenna_layout_aware}
\end{figure}

For a $N_\mathrm{t}$-antenna UPA, its antennas are arranged in a plane in a column-major manner. The location of each antenna $p$ ($p=0,1,\cdots,N_\mathrm{t}-1$) can be represented as $\left(x_p,y_p\right)$ in an $M$-by-$N$ grid, where $M$ is the number of antennas in each column and $N=N_\mathrm{t}/M$ is the number of antennas in each row, with $0\leq x_p \leq N-1 $ and  $0\leq y_p \leq M-1$. An example of a $24$-antenna array arranged as a $4$-by-$6$ panel is depicted in Fig.~\ref{fig_Antenna_layout_aware}. Given the antenna index $p$, the coordinates $\left(x_p,y_p\right)$ can be calculated as
\begin{align}
x_p&=\lfloor p/M \rfloor\nonumber\\
y_p&= \mathrm{mod}\left(p,M\right)\nonumber
\end{align}
where $\lfloor x \rfloor$ is the maximum integer less than $x$ and $\mathrm{mod}\left(p,M\right)$ is the modulo operation with respect to $M$.
On the contrary, when the coordinates $\left(x_p,y_p\right)$ of an antenna are given, the antenna index $p$ can be computed as $p=y_pM+x_p$.

To describe the principles of ALA covariance matrix estimation for UPA, let us consider an antenna pair PAIR$\left(p,q \right)$. It can be noticed that, when an angular spectrum is provided, the covariance value of PAIR$\left(p,q \right)$ depends only on the relative positions of antennas $p$ and $q$. As a result, the covariance value of an antenna pair is translation-invariant in the 2D grid. An example is illustrated in Fig.~\ref{fig_Antenna_layout_aware}. Let us consider PAIR($p,q$) and PAIR($p',q'$). The antenna relative positions in these two pairs are the same and their covariance values are statistically equivalent, i.e., $\left[\mathbf{R} \right]_{pq}=\left[\mathbf{R} \right]_{p'q'}$. However, covariance values are not rotation-invariant to antenna pairs, because the angular spread in horizontal may not be the same as that in elevation. Rotating the antenna pair will change its covariance value. Then, the estimated covariance value can be calculated as the average of all covariance values of antenna pairs with the same relative positions.

Next, assuming a given PAIR($p,q$), the remaining question is how to identify the set $\mathcal{S}_{pq}$ consisting of all antenna pairs with the same relative antenna positions as PAIR($p,q$). PAIR($p',q'$) is a translation of PAIR($p,q$) if the coordinates in PAIR($p',q'$) can be expressed as shifted versions of coordinates in PAIR($p,q$). These can be presented as
\begin{align}
x_{p'}=x_{p}+\Delta x, 
x_{q'}=x_{q}+\Delta x
\end{align}
\begin{align}
y_{p'}=y_{p}+\Delta y, 
y_{q'}=y_{q}+\Delta y
\end{align}
Considering that the ranges of the horizontal direction and the vertical direction are bounded by $N$ and $M$,  the set $\mathcal{S}_{pq}$ can be expressed as
\begin{align}
\mathcal{S}_{pq}&=\left\lbrace  \left(x_{p}+\Delta x,y_{p}+\Delta y\right),\left(x_{q}+\Delta x,y_{q}+\Delta y\right)| \right.\nonumber\\
&\left. -\min \left\lbrace x_p,x_q \right\rbrace\leq \Delta x \leq N-1-\max \left\lbrace x_p,x_q \right\rbrace,\right.\nonumber\\
&\left. -\min \left\lbrace y_p,y_q \right\rbrace\leq \Delta y \leq M-1-\max \left\lbrace y_p,y_q \right\rbrace \right\rbrace.
\end{align}
Moreover, the cardinality $|\mathcal{S}_{pq}|$ can be calculated as 
\begin{align}
|\mathcal{S}_{pq}|=\left(N-|x_p-x_q|\right)\left(M-|y_p-y_q|\right).
\end{align}
As a result, elements in the estimated covariance matrix using ALA can be computed as
\begin{align}
\left[\hat{\mathbf{Q}} \right]_{pq}=\frac{1}{|\mathcal{S}_{pq}|}\sum\limits_{\mathrm{PAIR}(p',q')\in |\mathcal{S}_{pq}| }\left[\hat{\mathbf{Q}}^{\mathrm{sample}} \right]_{p'q'}.
\end{align}
The pseudo codes of the ALA covariance matrix estimation for UPA are shown in Fig.~\ref{fig_ALA_2D}. It can be seen in Fig.~\ref{fig_ALA_2D} that a $N_\mathrm{t}$-by-$N_\mathrm{t}$ boolean assistance matrix is used to mark which elements have been calculated. If an element has been already calculated, the loop will be skipped. This assistance matrix can be used to avoid repeated computations, minimizing the complexity.

\begin{figure}
\hrulefill
\begin{algorithmic}[1]
\Require $\hat{\mathbf{Q}}^{\text{(sample)}}$

\State flag=false($N_\mathrm{t},N_\mathrm{t}$)
                \For{each PAIR($p,q$)}

                \If {flag($p,q$)==false}
                \State $c=0$;$b=0$
                \State $L_x$=-$\min \left\lbrace x_p,x_q \right\rbrace$;$R_x=N-1-\max \left\lbrace x_p,x_q \right\rbrace$
                \State $L_y$=-$\min \left\lbrace y_p,y_q \right\rbrace$;$R_y=M-1-\max \left\lbrace y_p,y_q \right\rbrace$

                \For {$\Delta x$=$L_x$:$R_x$}
               
                \For {$\Delta y$=$L_y$:$R_y$}
                 \State $p'=(x_p+\Delta x)M+(y_p+\Delta y)$
                \State $q'=(x_q+\Delta x)M+(y_q+\Delta y)$
                \State $b=b+\hat{\mathbf{Q}}^{\text{(sample)}}\left(p',q' \right)$
                \State $c=c+1$

                \EndFor
                \EndFor
                \For {$\Delta x$=$L_x$:$R_x$}
               
                \For {$\Delta y$=$L_y$:$R_y$}
                 \State $p'=(x_p+\Delta x)M+(y_p+\Delta y)$
                \State $q'=(x_q+\Delta x)M+(y_q+\Delta y)$
               
                \State $\hat{\mathbf{Q}}\left(p',q' \right)=b/c$;$\hat{\mathbf{Q}}\left(q',p' \right)=\mathrm{conj}(b)/c$
                \State flag$\left(p',q' \right)=\mathrm{true}$;flag$\left(q',p' \right)=\mathrm{true}$
                
                \EndFor
                \EndFor
                \EndIf

                \EndFor

\end{algorithmic}

\hrulefill
\caption{ALA covariance matrix estimation for UPA.}
\label{fig_ALA_2D}
\end{figure}

\subsection{General antenna layouts}
Although the most two typical antenna layouts, i.e., ULA and UPA, are discussed in this paper, the proposed ALA covariance matrix estimation algorithm can be applied to more general antenna layouts as long as the following two conditions are satisfied. First, the antenna layout information needs to be provided to the estimator as prior information. Second, only antenna pairs with translation operations can contribute when calculating the average covariance value. 
\subsection{Complexity analysis of the ALA estimation method}
When the number of antennas is relatively large, maintaining low complexity becomes crucial. It can be observed in Fig.~\ref{fig_ALA_1D} and Fig.~\ref{fig_ALA_2D}  that each element in the covariance matrix is visited only once. Therefore, the ALA estimation method has polynomial complexity $O(N^2_{\mathrm{t}})$ .

\section{Results and Analysis} \label{sec_results_section}
Simulations are performed in a one-ring seven-cell network. Uplink power settings are assumed to be the same as \cite{Bjornson16}, i.e., $\rho^{\rm UL}_{lk}= -7\mathrm{dB}$ if $l=k$ and $\rho^{\rm UL}_{lk}= -8.6\mathrm{dB}$ if $l\neq k$. In downlink, the SNR of the attached UE is $13$ dB and the SNRs of UEs in neighbor cells are calculated proportionally to their distances to the the BS assuming the path loss exponent is $2$. For example, in Fig.~\ref{fig_System_diagram}, let us assume that the distance between the  BS$1$ and its attached UE (center UE) is $D$ and the downlink SNR is $13$ dB. Then, the distance between BS2 and the center UE is $2D$. With path loss exponent $2$, it can be obtained that the SNR between BS2 and the center UE is $7$ dB. All other SNRs are computed in the same way. Moreover, correlation factors are set as $r_{\mathrm{h}}=0.5$ and $r_{\mathrm{v}}=0.65$ for azimuth and elevation. Azimuth AoAs $\theta_{lk,h}$ are drawn from a uniform distribution within $\left(-\pi,\pi \right)$ and elevation AoAs $\theta_{lk,v}$ are drawn from a uniform distribution within $\left(-\pi/2,\pi/2 \right)$.

\begin{figure} 
\centering\includegraphics[width=3.5in]{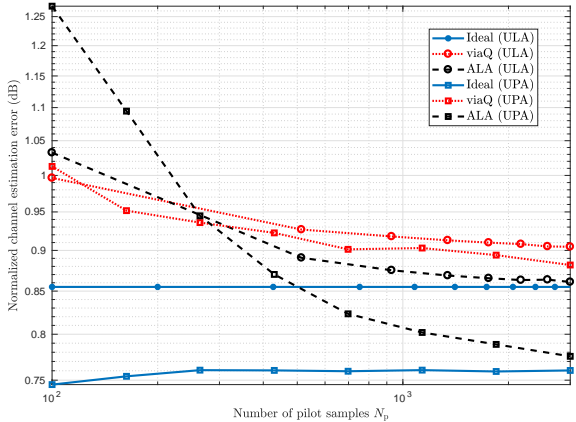} 
\caption{Channel estimation MSE with respect to different covariance matrix estimation methods, number of pilot samples, and antenna layouts ($N_\mathrm{t}=128$).}
\label{fig_mse}
\end{figure}

The normalized channel estimation errors with different covariance matrix estimation methods, number of pilot samples, and antenna layouts are depicted in Fig.~\ref{fig_mse}. The normalized channel estimation error is measured by the mean squared error (MSE) and is calculated as \cite{Bjornson16}
\begin{align}
\mathrm{MSE}=\mathbb{E}\left\{\parallel\mathbf{g}_{lk}-\hat{\mathbf{g}}_{lk}\parallel^2\right\}/\mathrm{tr}(\mathbf{R}_{lk})
\end{align}
The MSE curves with ideal covariance matrix serve as lower bounds and remain flat  in terms of $N_\mathrm{p}$ as expected. When $N_\mathrm{p}$ is relatively small ($N_\mathrm{p}\leq 250$), the viaQ method outperforms the proposed ALA method.  The reason for this is that multiple entries in the estimated covariance matrix of the ALA method share a common value. When this common value is not accurate enough due to insufficient number of pilots, the estimated error propagates, causing the slightly poorer performance in the ALA method when $N_\mathrm{p}$ is relatively small. However, when $N_\mathrm{p}$ continues to grow, the proposed ALA method dominates and is able to close its gap to the ideal covariance matrix curve more quickly than the viaQ method. The viaQ method suffers from its nature in (\ref{equ_viaQ_regularization}) that its off-diagonal entries are always scaled and $1-\kappa$ times smaller than those in the sample covariance matrix. The impact of these scaled entries is shown to be significant when $N_\mathrm{p}$ is sufficiently large.

\begin{figure} 
\centering\includegraphics[width=3.5in]{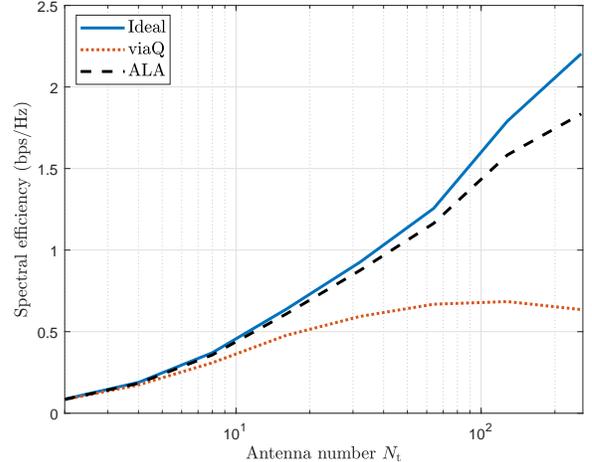} 
\caption{Downlink spectral efficiency with ULA ($N_\mathrm{p}=3000$).}
\label{fig_SE_1D}
\end{figure}

Downlink spectral efficiency with ULA is illustrated in Fig.~\ref{fig_SE_1D}. The spectral efficiency with ideal covariance matrix serves as the upper bound and grows with the antenna number. When the antenna number is small, the ALA and viaQ achieve similar performance. However, when the antenna number is moderate and large $64\leq N_\mathrm{t}\leq 256$, the ALA method achieves better spectral efficiency than the viaQ method. It can also be observed that the spectral efficiency of the viaQ method starts to saturate, while the spectral efficiency of the ALA method continues to increase with antenna number.

\begin{figure} 
\centering\includegraphics[width=3.5in]{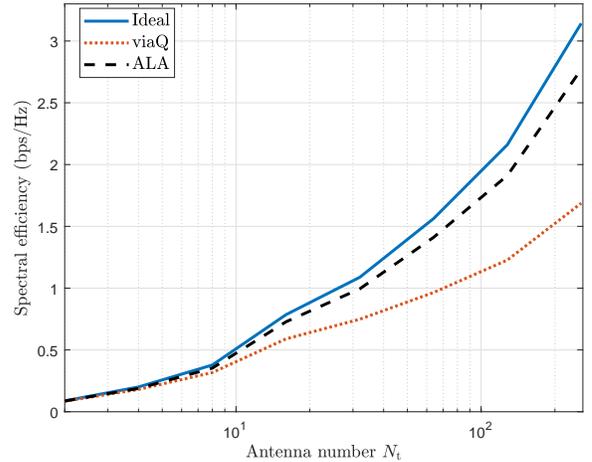} 
\caption{Downlink spectral efficiency with UPA ($N_\mathrm{p}=3000$).}
\label{fig_SE_2D}
\end{figure}

Downlink spectral efficiency with UPA is shown in Fig.~\ref{fig_SE_2D}. Spectral efficiencies of both the ideal and ALA method show a similar trend as those using ULA. What is more, the spectral efficiency of the viaQ method is able to grow with the antenna number as well. This shows that UPA is more resilient to covariance matrix estimation errors than ULA. The reason for this can be explained in Table \ref{tab_kappa}. It can be seen that the viaQ $\kappa$ for ULA converges to $1$ after 128 antennas, meaning that the viaQ estimated covariance matrix is fully determined by the diagonal matrix of the sample covariance matrix. On the other hand, the viaQ $\kappa$ for UPA converges much slower. As a result, the off-diagonal entries of the sample covariance matrix still contribute to the estimated covariance matrix.

\begin{table}[ht]
\caption{Values of the viaQ regularization factor $\kappa$ in different antenna layouts and numbers ($N_\mathrm{p}=3000$).}
\center

    \begin{tabular}{|c|c|c|c|c|c|c|c|c|}
    \hline
      & $N_\mathrm{t}=2$ & 4 & 8 & 16 & 32 & 64 & 128 & 256 \\ \hline
ULA  & 0.29 & 0.38 & 0.54 & 0.70 & 0.84 & 0.96 & 1.00 & 1.00   \\ \hline        
        UPA  & 0.29   & 0.26   & 0.37 & 0.45 & 0.60 & 0.71 & 0.84 & 0.93  \\ \hline

    \end{tabular}
    \label{tab_kappa}
\end{table}

\section{Conclusions} \label{sec_conclusion_section}
A low-complexity ALA covariance matrix estimation method has been presented in this paper. Since covariance matrices do not have full degrees of freedom, the proposed ALA method can maximize the benefit of this property by allowing a BS to have knowledge of its antenna layout. It has been shown that the ALA method is effective to both ULA and UPA layouts, which are the two most widely used in practice. The proposed ALA method has been applied to a multi-cell network. Simulations have shown that the proposed ALA method has lower MSE of the estimated channel than the viaQ method when the number of pilot symbols is moderate. Additionally, the proposed ALA method significantly outperforms the viaQ method in terms of spectral efficiency. For future work, the application of the ALA method to multi-cell multi-user network in both uplink and downlink can be investigated. Moreover, impact of effects such as antenna rotations, misalignment, and imperfect knowledge of antenna geometry can be studied.


\section*{Acknowledgment}
Part of this work has been performed in the framework of the Horizon 2020 project ONE5G (ICT-760809) receiving funds from the European Union. The authors would like to acknowledge the contributions of their colleagues in the project, although the views expressed in this contribution are those of the authors and do not necessarily represent the project.


%



%

\ifCLASSOPTIONcaptionsoff
  \newpage
\fi

\end{document}